\documentclass[journal=ancac3,manuscript=article]{achemso}

\usepackage[T1]{fontenc}
\usepackage{lmodern}

\usepackage{microtype}
\usepackage{xurl}
\usepackage{xcolor}

\makeatletter
\expandafter\let\csname ver@natbib.sty\endcsname\relax
\expandafter\let\csname ver@mciteplus.sty\endcsname\relax
   \@for\acs@tmp:={citet,citep,citealt,citealp,Citet,Citep,Citealt,Citealp,citeauthor,Citeauthor,citeyear,citeyearpar,citetext,citenum,citename,citefullauthor,Citefullauthor,bibfont,bibhang,bibsep,bibpunct,citestyle,defcitealias,citetalias,citepalias,citenumfont,bibnumfmt}\do{%
\expandafter\let\csname\acs@tmp\endcsname\@undefined}
\makeatother
\usepackage[style = chem-acs, articletitle = true, doi = true, maxnames = 99]{biblatex}
\usepackage{graphicx}
\graphicspath{{figures/}}

\usepackage[version = 4]{mhchem} 
\usepackage{amsmath}
\usepackage{bm}                  

\author{Swarnendu Das}
\altaffiliation{These authors contributed equally to this work.}
\affiliation{Department of Materials Science and Engineering, University of Pennsylvania, Philadelphia, Pennsylvania, 19104, United States}
 
\author{Shengsong Yang}
\altaffiliation{These authors contributed equally to this work.}
\affiliation{Department of Chemistry, University of Pennsylvania, Philadelphia, Pennsylvania, 19104, United States}
 
\author{Kevin N. Moser}
\author{Marc R. Bourgeois}
\affiliation{Department of Chemistry, University of Washington, Seattle, Washington, 98195, United States}
 
\author{Quentin M. Ramasse}
\affiliation{SuperSTEM, SciTech Daresbury Science and Innovation Campus, Block J, Keckwick Lane, Daresbury WA4 4AD, United Kingdom}
\alsoaffiliation{School of Chemical and Process Engineering and School of Physics and Astronomy, University of Leeds, Leeds LS2 9JT, United Kingdom}
 
\author{David J. Masiello}
\affiliation{Department of Chemistry, University of Washington, Seattle, Washington, 98195, United States}
 
\author{Christopher B. Murray}
\email{cbmurray@sas.upenn.edu}
\affiliation{Department of Materials Science and Engineering, University of Pennsylvania, Philadelphia, Pennsylvania, 19104, United States}
\alsoaffiliation{Department of Chemistry, University of Pennsylvania, Philadelphia, Pennsylvania, 19104, United States}
 
\author{Eric A. Stach}
\email{stach@seas.upenn.edu}
\affiliation{Department of Materials Science and Engineering, University of Pennsylvania, Philadelphia, Pennsylvania, 19104, United States}

\title{Probing Plasmonic Oscillations in 2D Moir\'e Nanocrystal
  Superlattices by Low-Loss EELS}


\begin{document}

\begin{tocentry}
\includegraphics[width=3.25in,height=1.75in,keepaspectratio]{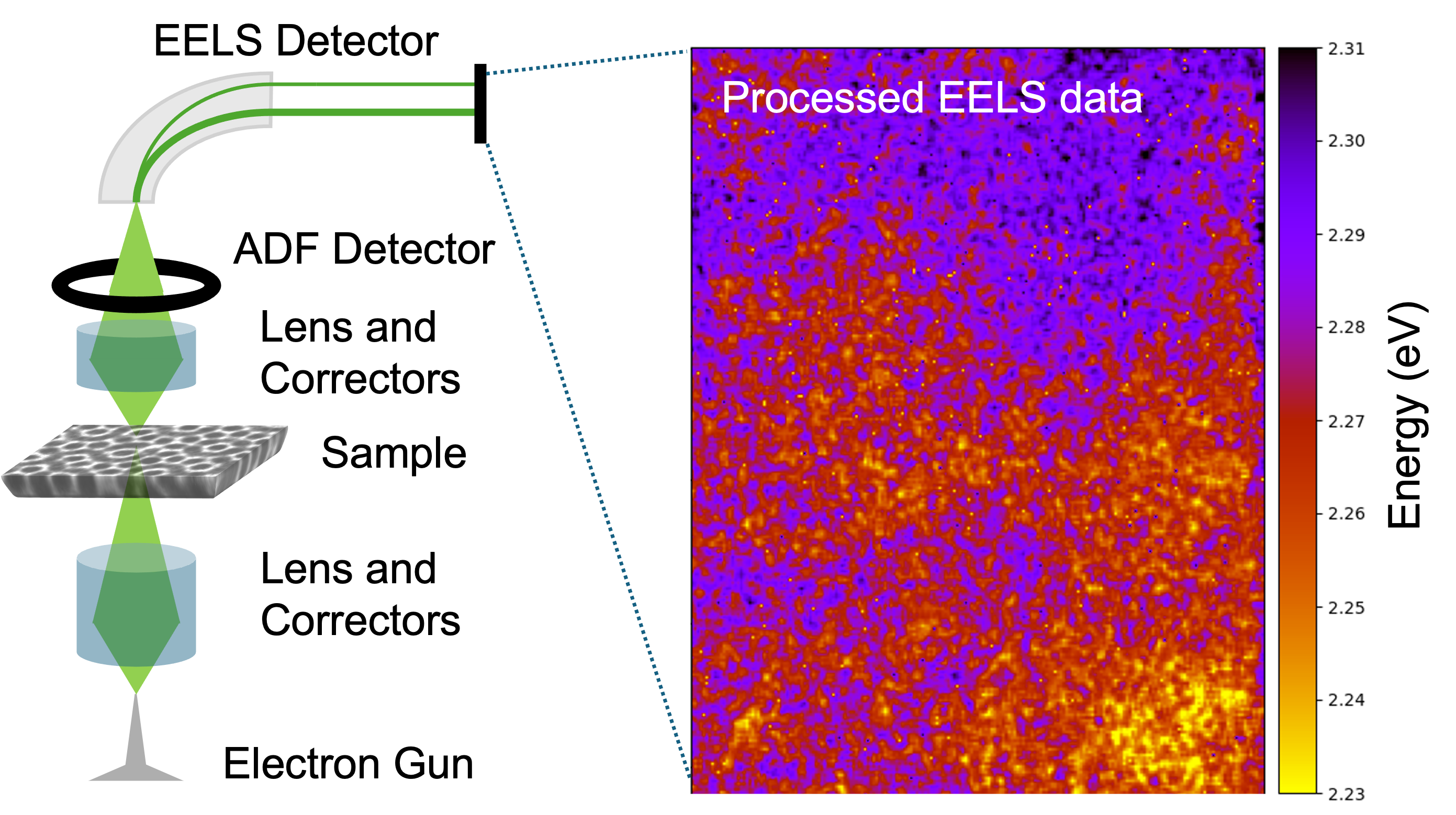}
\end{tocentry}

\begin{abstract}
Electron energy loss spectroscopy (EELS) has been established as a powerful analytical technique for investigating the oxidation state, band structure, and dielectric properties of materials with exceptional spatial resolution. Inspired by twisted 2D materials, we utilize low-loss EELS to examine the plasmonic excitations in 2D moir\'e Au nanocrystal superlattices (NCSLs) formed by liquid-air interface self-assembly using a double-dipping method. This approach produces stacked hexagonal layers that can be twisted, forming moir\'e patterns in NCSLs whose twist angles are precisely measured via scanning transmission electron microscopy (STEM). Low-loss EELS effectively mitigates challenges arising from fabrication-induced non-uniformity and reveals a blue shift in plasmonic excitation when comparing monolayer, bilayer, and twisted configurations. This sharply contrasts with the optical spectroscopy measurements, which show an overall red shift relative to the EELS data. Coupled dipole modeling explains the observed discrepancies: the electron beam excites out-of-plane polarized lattice modes inaccessible to far-field optical radiation, while optical measurements average over ensembles. The high spatial resolution of STEM-EELS further demonstrates that twist-induced symmetry breaking locally influences the plasmonic behavior of these NCSL systems, unlike the twisted 2D materials. Our findings highlight that EELS provides complementary information to optical spectroscopy for understanding how structural arrangements at the nanoscale influence collective electronic properties, advancing the design of plasmonic metamaterials.
\end{abstract}

\section{Introduction}

Functionalized nanocrystals (NCs) have been realized as `artificial atoms',\autocite{murray1995,love2005} allowing the creation of novel materials through their assembly into superlattices (SLs). Nanocrystal superlattices (NCSLs) allow tuning of electrical,\autocite{cargnello2015,coropceanu2022} optical,\autocite{schulz2020,mueller2020} and mechanical properties.\autocite{dhulipala2023,lewis2020} The assembly of plasmonic nanoparticles into ordered 2D- and 3D-SLs has led to new tailored materials for plasmonic sensing, photocatalysis, and manipulation of light on the nanoscale.\autocite{schulz2020,mueller2020,ye2013} However, it is anticipated that the properties of such materials will strongly depend on the interplay between NC size, shape, composition, and ordering, in both two and three dimensions.

Moir\'e patterns are large-scale interference patterns that can be produced when a partially opaque ruled pattern with transparent gaps is overlaid on another similar pattern.\autocite{he2021,song2022,quan2021} These patterns appear in many instances, and from macroscopic to nanoscopic length scales. As a quasi-periodic geometric pattern generated by the incommensurate stacking between two monolayers, twisted van der Waals (vdW) 2D materials have attracted the most attention due to their unique ability to modulate electronic properties.\autocite{tang2020,moire2021} The quasi-periodic geometry can also greatly expand the structural diversity of crystalline materials towards quasi-crystallinity and more.

SLs assembled from plasmonic NCs demonstrate interparticle coupling that can be tuned through packing density and local ordering.\autocite{ye2013,schulz2020} Emerging applications of surface-enhanced Raman spectroscopy and thermoplasmonics seek a more precise control of the physical locations of plasmonic hotspots, wavelength, and the strength of resonance.\autocite{sharma2025} Consequently, fabricating diverse nanocrystal superlattices enables a wide range of programmable plasmonic patterns. Different approaches including ligand design and controlled evaporation have been applied during NCSL formation, and moir\'e patterns can be occasionally observed.\autocite{schulz2020,dong2011,jishkariani2019} Analogous to vdW materials systems, deliberate stacking of two well-ordered monolayer NCSLs represents a robust route to engineered 2D moir\'e NCSLs and is compatible with liquid--air interfacial self-assembly.\autocite{dong2011} However, limitations in grain size and heterogeneity in thickness, twist angle, and grain boundaries complicate optical spectroscopy measurements, hindering the establishment of precise structure--property relationships.

Electron energy loss spectroscopy (EELS) is an analytical technique that is based on inelastic scattering of fast electrons in a thin specimen.\autocite{hofer2016} EELS has become a powerful tool for investigating the oxidation state,\autocite{foucher2023,foucher2022} band structure,\autocite{song2022,zhan2018} and dielectric properties of materials at the nanoscale.\autocite{brokkelkamp2022} When implemented in an advanced monochromated, aberration-corrected scanning transmission electron microscope (STEM), EELS provides information with both high spatial and energy resolution. The low-loss or valence region ($<50$ eV) of an EEL spectrum provides complementary information to that provided by optical spectroscopy, and contains valuable information about valence electron oscillations, such as plasmonic excitations.\autocite{koh2009,wu2018,zhou2022} The high spatial resolution of aberration-corrected STEM enables nanometer-scale imaging and identification of regions with different moir\'e orientations (Fig. S2). Furthermore, with the high energy resolution of monochromated EELS, the plasmonic response of neighboring moir\'e domains can be probed. Thus, STEM-EELS is capable of correlating structural information to collective plasmonic excitations.

In this work, we demonstrate the limitations of optical spectroscopy techniques for studying plasmonic metamaterials with fabrication-induced non-uniformity. While STEM-EELS is successful in mitigating these challenges, there are differences between local EELS and bulk optical responses caused by differences in their interaction physics. We clarify these differences by modeling the dynamics of plasmonic excitations in NCSLs using the method of coupled-dipole under both optical plane wave and electron beam illumination conditions.

\section{Results and Discussion}

\subsection{Fabrication of Moir\'e Au NCSLs}

\begin{figure}[H]
  \centering
  \includegraphics[width=0.78\textwidth]{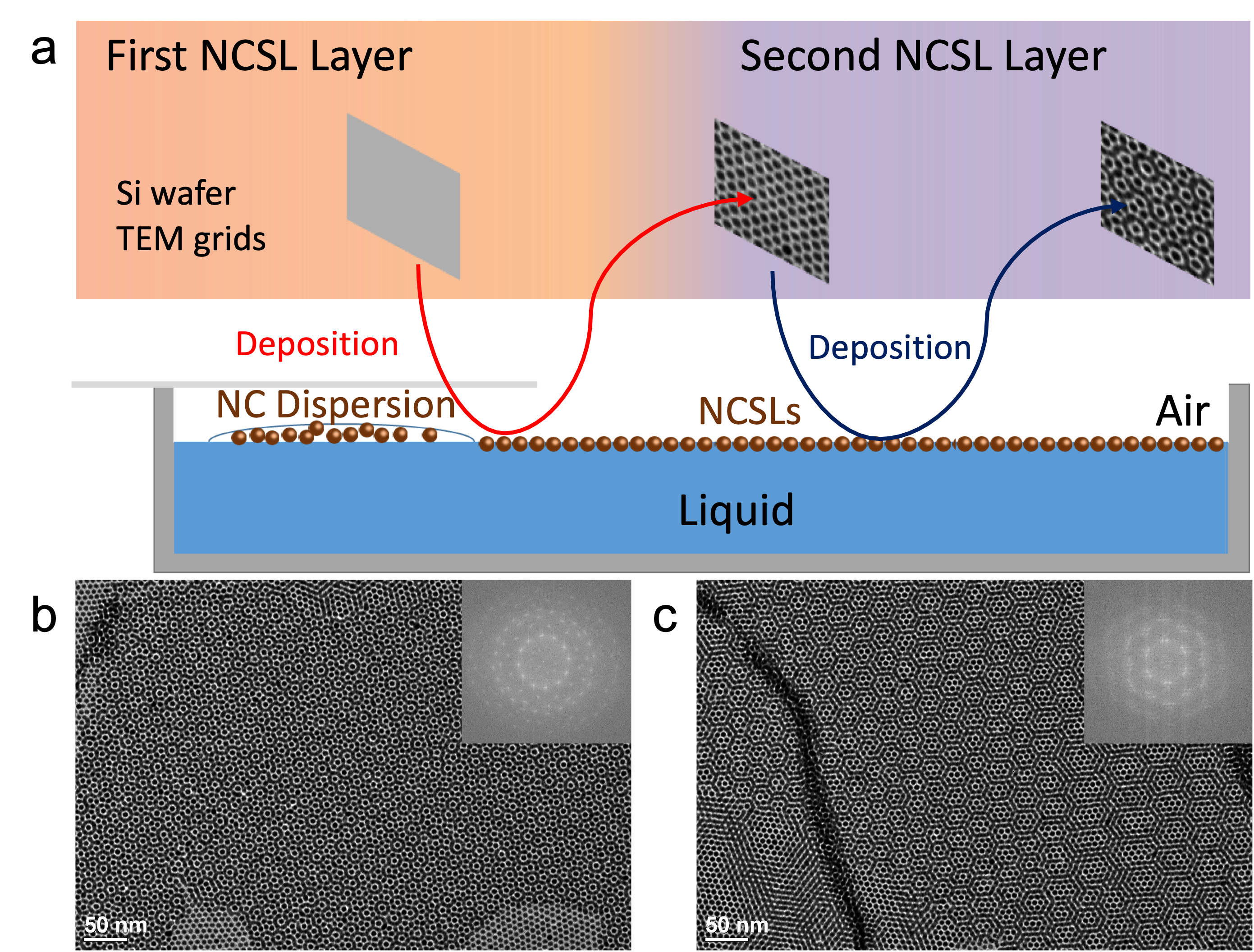}
  \caption{(a) Schematics of double deposition of self-assembled layers at the liquid-air interface. (b) Au NCSL twisted bilayer showing dodecagonal quasicrystal ordering. (c) Au NCSL twisted bilayer showing moir\'e pattern from a different twist angle (less than 30\textdegree). Grain boundaries can be seen. Insets are fast Fourier transforms (FFTs), demonstrating the ordering.}
  \label{fig:1}
\end{figure}

High-quality NCSL monolayers can be produced by self-assembly at the liquid-air interface.\autocite{dong2011} This method can be repeated to deposit multiple layers of the same or different NCSLs on one substrate to generate 2D moir\'e patterned NCSLs (Fig.~\ref{fig:1}a). Here, we use 5~nm Au NCs for the hexagonal close-packed (HCP) monolayers and a double-dipping method.

This double-dipping method generally lacks control of the twisting angle in the microstructures and yields many different moir\'e patterns, such as dodecagonal quasicrystals (DDQC) (Fig.~\ref{fig:1}b), strips, and hexagons (Fig.~\ref{fig:1}c). Furthermore, the SL films may be broken during transfer and deposition. These factors make it challenging to create a uniform and homogeneous moir\'e pattern over several micrometers, which is critical for microscopic optical characterizations.

\subsection{Optical Microspectroscopy}

\begin{figure}[H]
  \centering
  \includegraphics[width=0.7\textwidth]{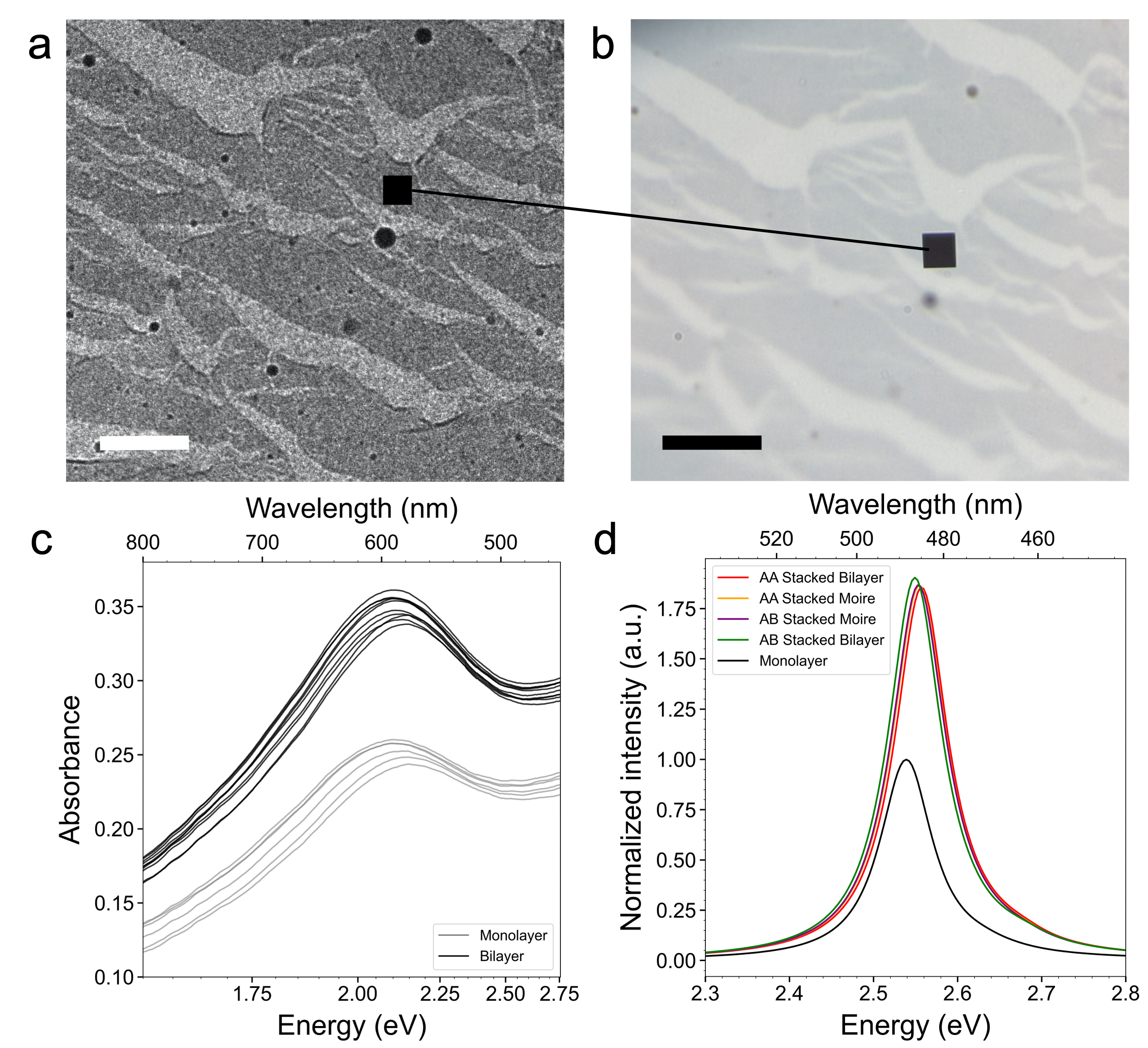}
  \caption{(a) A low-magnification TEM image of our NCSL sample. (b) The same region imaged using optical microscopy. The black box shows a representative 3.1$\times$3.1 $\mu$m region from which we obtained absorption spectra. (c) Absorption spectroscopy of several different regions captured from the NCSL film shown in (a) and (b). (d) Model calculation of optical extinction spectra for different numbers of NCSL layers and stacking configurations. Scale bars in (a) and (b): 10 $\mu$m.}
  \label{fig:2}
\end{figure}

Optical microspectroscopy was used to provide initial information on the optical response of our Au NCSL samples. Optical microspectroscopy can be used to capture absorption spectra of a region as small as 3.1$\times$3.1 $\mu$m\autocite{ye2013}. However, this size is larger than the average grain size of a single correlated moir\'e region. Multiple regions of monolayer and bilayer NCSLs with different moir\'e configurations were identified by transmission electron microscopy (TEM), and the optical absorption spectra from each region was collected. Monolayer and bilayer regions were initially distinguished by the contrast and later confirmed by imaging at higher magnification (Fig.~S1). A Gilder finder grid\autocite{gilder} was used to correlate the regions between TEM and optical microscopy (Fig.~\ref{fig:2}a--b). The absorbance data in Fig.~\ref{fig:2}c show broad plasmonic peaks centered at $585 \pm 5$~nm (2.12~eV) for both monolayer and bilayer regions, with bilayer NCSLs having higher absorbance than monolayers. Model calculation of the NCSL optical extinction spectra using the method of coupled dipoles (Methods) shows that optical stimulation is dependent on the number of layers, but less sensitive to the stacking sequence and twisting of the bilayers (Fig.~\ref{fig:2}d). The intensities are normalized by the monolayer spectrum intensity. Because of negligible contribution from scattering, the calculated optical extinction spectra can be compared with the experimental absorption spectra (Fig.~S11). The ensemble average nature of optical spectroscopy, coupled with the heterogeneity and small grain size of our 2D NCSL sample, results in the convolution of any subtle peak shifts into a broad plasmonic peak. The limitations of optical measurements compelled our subsequent low-loss EELS studies.

\subsection{Electron Energy Loss Spectroscopy}

Fig.~S2 shows a high-angle annular-dark-field scanning transmission electron microscopy (HAADF-STEM) image of a region in our 2D moir\'e Au NCSL sample. The EELS data for our study were acquired in a dedicated monochromated STEM, over an energy range of 0--10~eV. The spectral range can be divided into zero-loss, low-loss, and extended-loss regions which represent the main primary beam, inelastic scattering from plasmonic excitations in the Au NCs, and inelastic scattering from interband transitions of Au, respectively. In the following, we focus on the lower loss region from 2~eV to 3~eV which contains the effects of plasmonic excitations in the NCSLs.

For our EELS data, we selected a monochromator slit width that provided an adequate balance of energy resolution and beam current, i.e., one that maintains sufficient electrons in the probe to suppress noise, without limiting the line width of the plasmonic modes. HyperSpy, an open-source Python package\autocite{hyperspy}, was used for processing the data and generating the maps.

\begin{figure}[H]
  \centering
  \includegraphics[width=\textwidth]{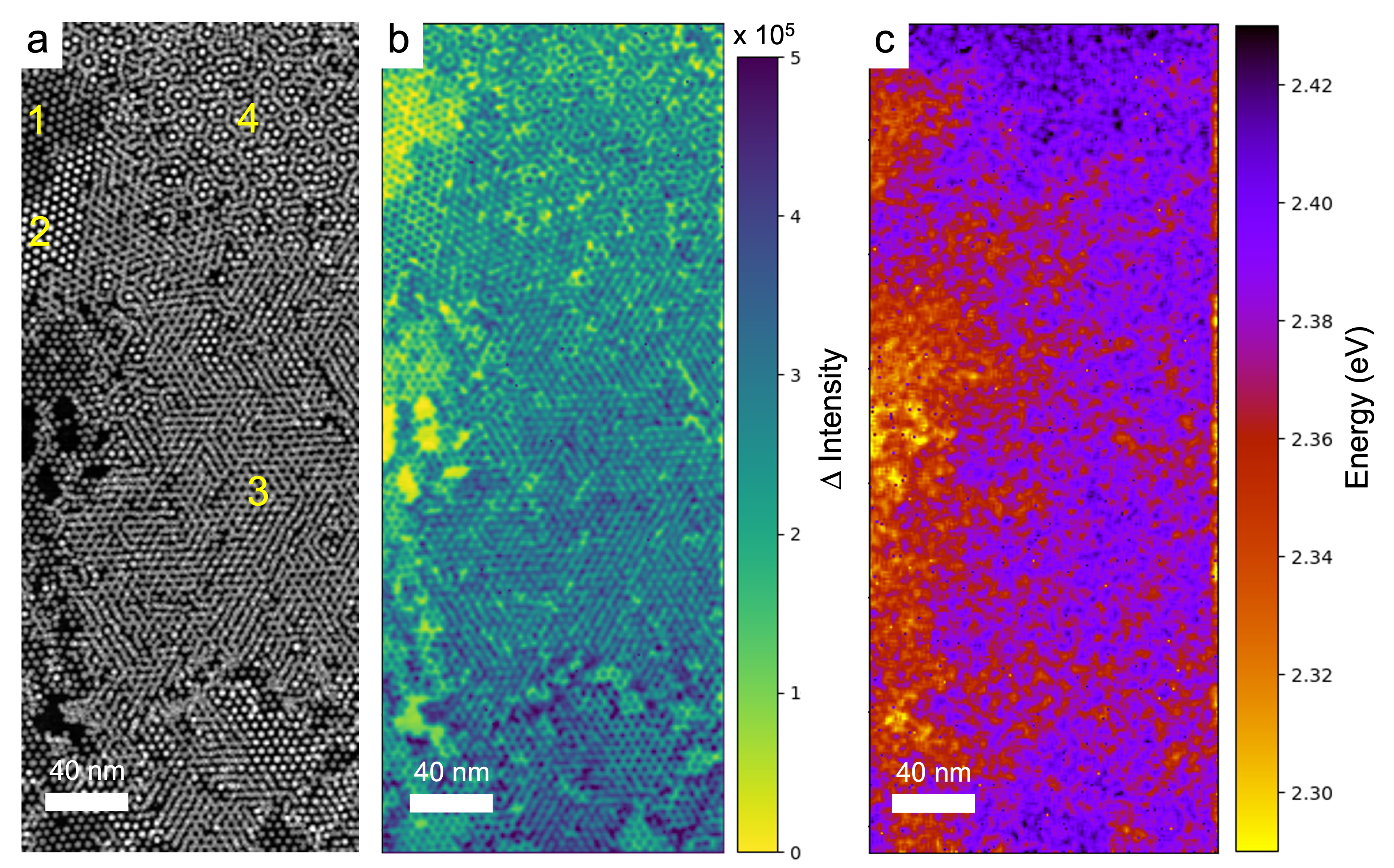}
  \caption{Effect of thickness on plasmonic excitations. (a) HAADF-STEM image of a multi-layered Au NCSLs with different thickness and twist angles. (b) Thickness profile extracted by the ZLP analysis from ROIs of $3\times3$ pixels. (c) Peak position of the plasmonic excitations from ROIs of $3\times3$ pixels.}
  \label{fig:3}
\end{figure}

\begin{figure}[H]
  \centering
  \includegraphics[width=\textwidth]{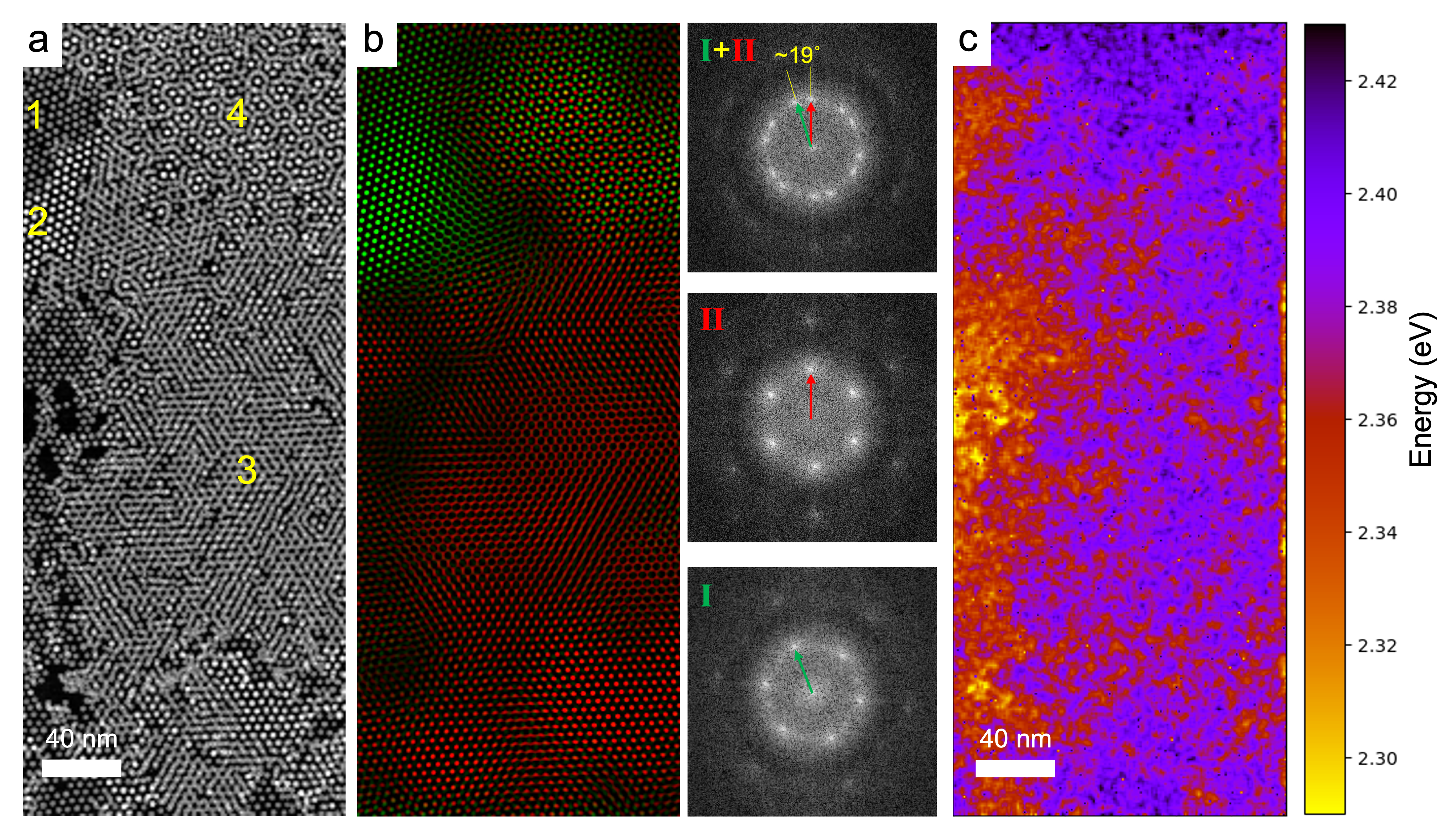}
  \caption{Effect of twist angle on plasmonic excitations. (a) HAADF-STEM image of a multi-layered Au NCSLs with different thickness and twist angles. (b) Inverse-FFT analysis to identify regions with different orientations. (c) Peak position of the plasmonic excitations from ROIs of $3\times3$ pixels.}
  \label{fig:4}
\end{figure}

\begin{figure}[H]
  \centering
  \includegraphics[width=\textwidth]{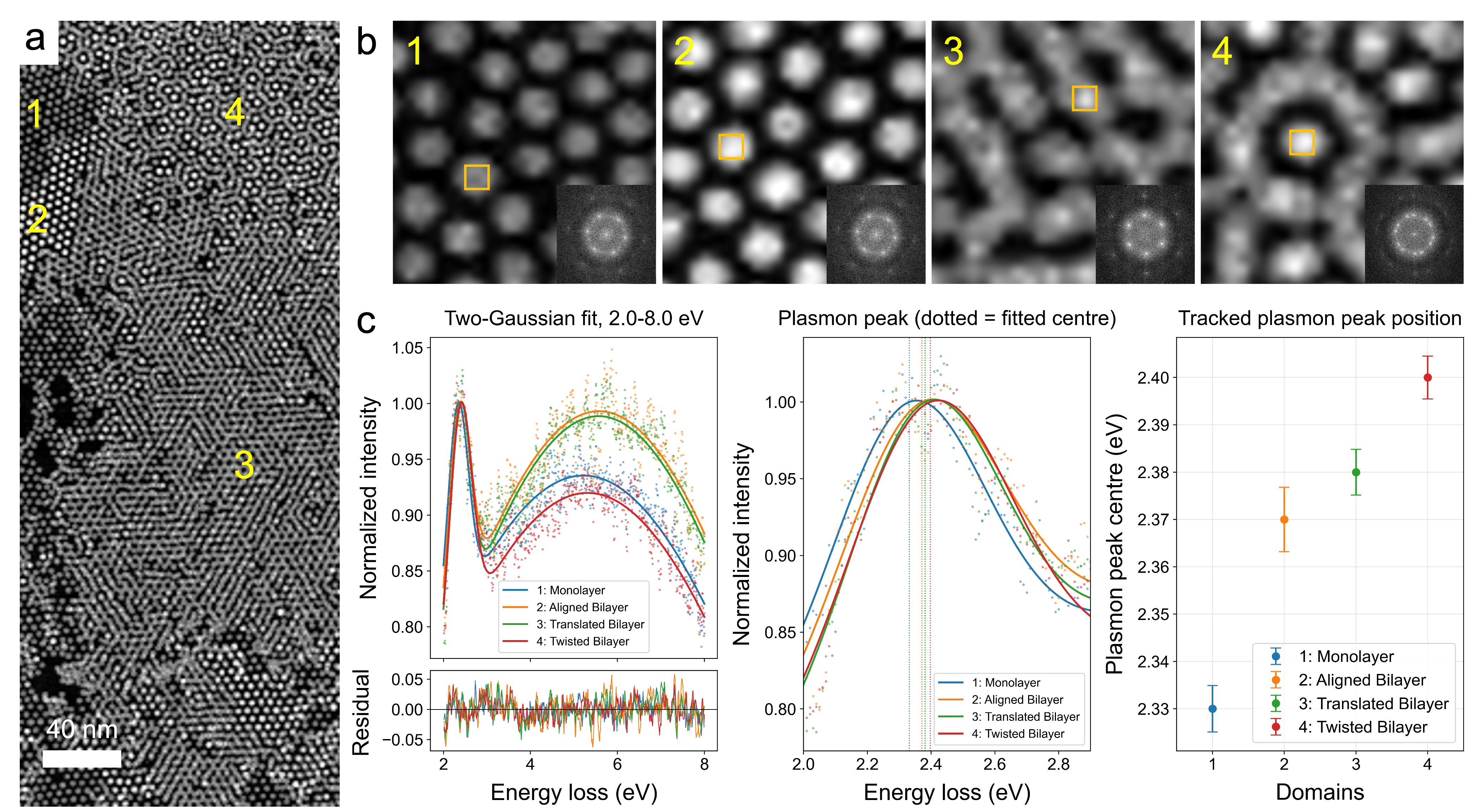}
  \caption{Representative EEL spectra from the four domains of interest: monolayer, aligned bilayer, translated bilayer and twisted bilayer, marked as 1, 2, 3 and 4 respectively. (a) HAADF-STEM image of a multi-layered Au NCSLs with different thickness and twist angles. (b) Magnified images of the four domains with the marked ROIs of $3\times3$ pixels (inset are the FFTs). (c) Raw and fitted EEL spectra from the four ROIs (left), zoomed plasmon peak showing the blue shift across different domains (middle), and the summary plot (right).}
  \label{fig:5}
\end{figure}

We have chosen four domains in our Au NCSL sample; monolayer, aligned bilayer, translated bilayer and twisted bilayer, marked as 1, 2, 3 and 4 respectively in Figs.~\ref{fig:3}a, \ref{fig:4}a and \ref{fig:5}a. Fig.~S3 highlights these domains with the help of schematics. The contrast in the real-space images comes from the scattering of the sample and is sensitive to both the element's electron scattering factors and the relative alignment of layers with respect to each other. As a result, the observed contrast difference may not represent the thickness accurately. A pure zero-loss peak analysis is a more accurate way to quantify the thickness in this case (see Methods). Fig.~\ref{fig:3}b shows the thickness profile extracted from the aforementioned method, which clearly distinguishes the number of SL layers. The highest intensity in the zero-loss peak corresponds to the region with the least thickness. A map was created reflecting this highest intensity, where zero relative intensity relates to the least thickness or even vacuum. The corresponding map indicates the thickness difference at multiple regions as the difference in intensity. As we can clearly see, the left part of the image contains several empty regions, i.e.\ regions with only an ultra-thin carbon support and devoid of any Au NP. There are also several regions in the bottom section of the image where there are three layers. The variety of patterns in this area allows investigation of the effect of structural diversity on electromagnetic response: this is the reason we chose this region for both data acquisition and subsequent analysis.

We extracted plasmonic excitation data from the EEL spectra. The combination of small pixel size and small pixel time results in low signal per pixel (see Methods). The data processing was optimized to balance signal quality and spatial resolution by implementing a rolling average using a region of interest-based scanning. When the region of interest (ROI) was insufficiently large, it resulted in a poor signal-to-noise ratio in the summed spectra, making curve fitting difficult. Conversely, if the ROI dimension was too large, the rolling average would smooth out energy shifts, preventing the detection of subtle spectral changes. After carefully balancing spatial resolution against signal-to-noise requirements, an ROI of 3 by 3 pixels was selected. The spectra from all nine pixels were summed to achieve a sufficient signal-to-noise ratio for accurate fitting (Fig.~S4 and~S5). As for the plasmonic signals, the summed spectrum from the ROI was fitted with two Gaussian curves for the two peaks within 2--8~eV. The peak positions and area under the curve were recorded from the fitted Gaussian curves corresponding to one particular ROI. The narrower peak at around 2.3~eV yields a map of the plasmonic peak position, as shown in Fig.~\ref{fig:3}c. Fig.~\ref{fig:3}c shows that the plasmonic excitation energy increased from 2.33~eV to 2.42~eV when comparing monolayer to bilayer structures. Fig.~S6 shows the histogram providing quantitative uncertainty analysis for the curve fitting. Based on our theoretical modeling (described later), plasmonic hybridization between Au NCSL layers results in this blue shift.

Based on the thickness information from the EELS data (Fig.~\ref{fig:3}b), there are monolayer, bilayer and even tri-layer domains in the region of the sample under investigation. In the bilayer structures with identifiable stacking variations, marked with yellow numbers in Fig.~\ref{fig:3}a, we observed plasmonic excitation shifts to higher energies (blue shift). While the highest blue shift (from 2.33~eV to 2.42~eV) corresponds to twist between the bilayers, there are two intermediate blue shifts, from 2.33~eV for monolayer to 2.38~eV for aligned bilayer, and then to 2.39~eV due to translation between the bilayers (Fig.~\ref{fig:3}c).

To identify domains with different orientations, we masked specific spots in the fast Fourier transform (FFT) of the HAADF-STEM image shown in Fig.~\ref{fig:4}a, and then used an inverse-FFT to identify the specific regions correlating to the chosen masks. Based on this analysis, there are only two domains in the region of the sample under investigation. The FFTs show that these two domains have the same six-fold symmetry, corresponding to the HCP structure of the self-assembled layers, but have different orientations with respect to each other. Fig.~\ref{fig:4}b shows the first orientation in green and the second one in red. In the top-right region in Fig.~\ref{fig:4}a where these two domains overlap (marked with number 4 in yellow), a distinct moir\'e pattern is noticeable. The twisting angle corresponding to this moir\'e pattern was calculated from its FFT (see Fig.~\ref{fig:4}b) and is approximately 19\textdegree. Correlating the inverse-FFT analysis with the corresponding plasmonic excitation map (Fig.~\ref{fig:4}c) confirms that the blue shift in plasmonic excitation from 2.33~eV to 2.42~eV is related to the 19\textdegree\ twist between the Au NCSL bilayers.

To correlate the effect of thickness and twist angle with the blue shift in plasmonic response, we studied the representative spectra from all four domains (as summarized in Fig.~\ref{fig:5}). The error bar is the standard error on the fitted center of the gaussian, taken from the fit's covariance matrix. It is worth noting that the extent of blue shift shown in Fig.~\ref{fig:5}c is lower than the values we reported earlier in the text by estimating visually from Fig.~\ref{fig:3}c and Fig.~\ref{fig:4}c. The values reported in Fig.~\ref{fig:5}c can be subjected to human bias because of manually selecting the ROIs. Since the overall trend remains unchanged, the exact values do not affect the conclusions.  It can be concluded that the blue shift from monolayer to aligned bilayer is purely due to thickness. Though the thickness decreases from aligned bilayer to translated bilayer, there is still a small blue shift. To understand the effect of twisting between the bilayers, we chose ROIs from the aligned bilayer and twisted bilayer domain such that the NCs are aligned vertically (Fig.~\ref{fig:5}b), thus keeping the thickness constant. For the twisted bilayer, the NCs are aligned perfectly along the twist axis. We will be calling these NCs `twist centers' in the following text. The plasmonic responses from the aligned NCs show a blue shift from aligned bilayer to the twisted one. Initially we thought a twist between two stacked layers can introduce an energy shift or the emergence of a new mode in plasmonic excitations.\autocite{schulz2020,mueller2020,kuang2024} But when we repeated our analysis with larger ROIs of $15\times15$ pixels to include multiple NCs (see Fig.~S7), we did not see noticeable difference in plasmonic response among the domains of different bilayer configuration. This indicates that, unlike the twisted vdW 2D materials systems, the blue shift due to twist between the NCSL bilayers is localized only at the twist centers (see Fig.~S8). This observation can be explained by the non-rigid nature of our NCSL monolayers. When the top monolayer is twisted, the NCs around the twist centers relax into the voids of the bottom monolayer. Unlike a rigid structure, this relaxation increases the number of nearest neighbors only for the twist center NCs, thus leading to stronger coupling and a blue shift. Our theoretical modeling, described in the next section, supports this explanation.

According to optical microspectroscopy, the plasmonic peaks are centered at 2.12~eV. In contrast, our EELS data -- with plasmonic peaks ranging from 2.33 to 2.42~eV -- shows an overall blue shift relative to our optical measurements. Excitation from an electron does not follow the same selection rules as optical probing, and therefore, optically inactive modes can be readily seen in EELS.\autocite{garciadeabajo2010,bigelow2012} Point electron excitation from STEM-EELS and global excitation from optical spectroscopy may also accentuate different mode couplings in plasmonic NCs.\autocite{nicoletti2013} To completely understand these differences, we performed theoretical calculations using the method of coupled dipoles.

\subsection{Coupled Dipole Modeling}

\begin{figure}[H]
  \centering
  \includegraphics[width=\textwidth]{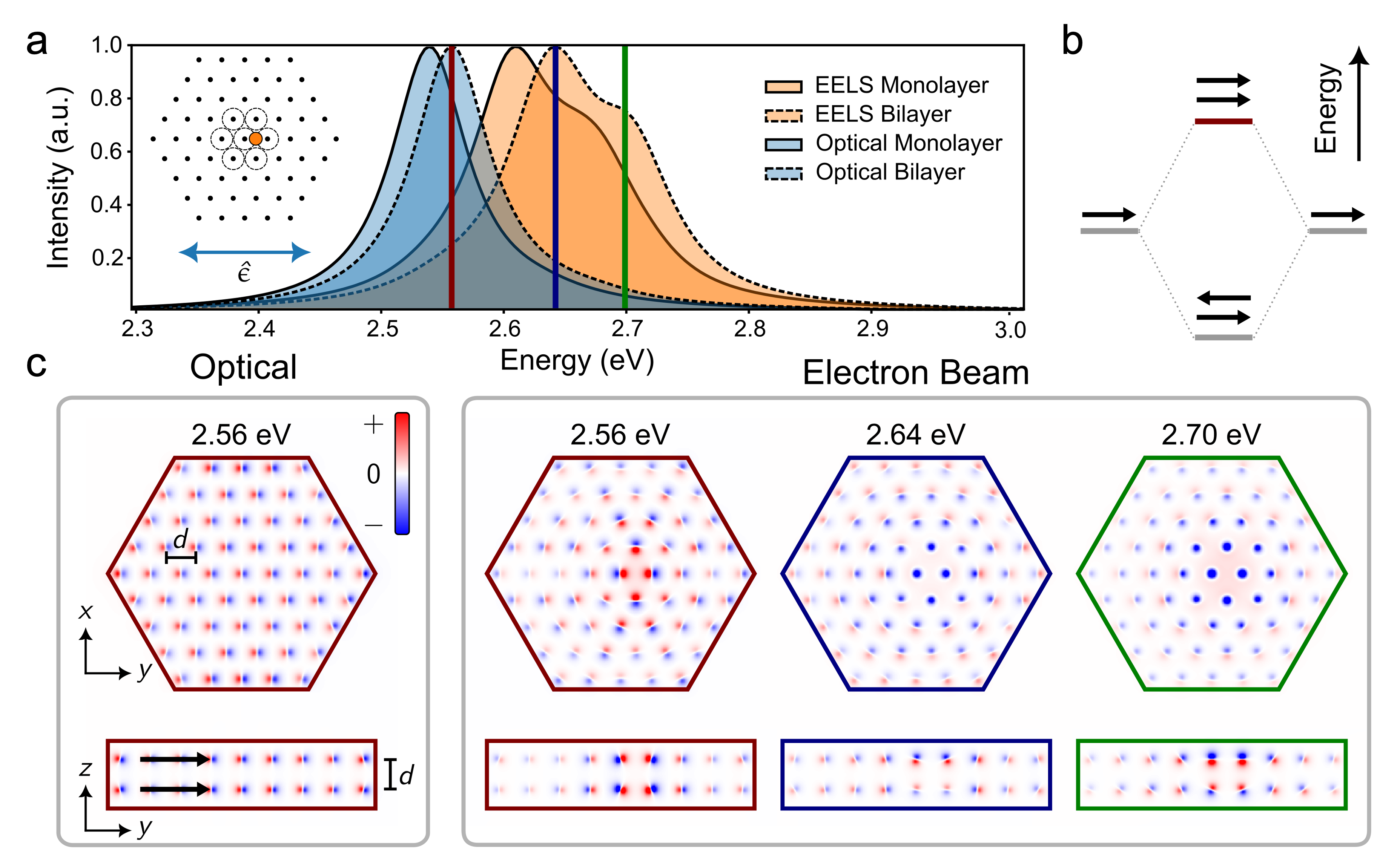}
  \caption{Model calculations of optical extinction and EEL spectra and induced polarization profiles at select loss energy values. (a) Optical extinction and EEL spectra for finite patches of both monolayer and bilayer hexagonal dipolar arrays, showing hybridization-induced energy shifts between the two array systems along with an apparent blue-shift in the EEL spectra. Also shown are the electron beam impact parameter (orange) and optical polarization vector $\hat{\bm{\epsilon}}$. Dashed lines indicate the boundaries of the nanospheres in their close-packed arrangement. (b) Hybridization diagram showing the in-plane out-of-phase (lower energy) and in-phase (higher energy) lattice modes of the bilayer NCSL. (c) The induced out-of-plane polarizations under optical (left) and electron beam (right) illumination of a bilayer hexagonal patch (at energies marked by vertical lines in (a)). In all cases, the STEM electron propagates along $\hat{\mathbf{z}}$ and the lattice spacing is indicated by $d$.}
  \label{fig:6}
\end{figure}
To elucidate trends observed in experimental optical absorption and EELS data, model calculations are performed on a finite hexagonal lattice patch with lattice spacing $d = 6.84$ nm, consistent with the experiment (see Methods). We note that model calculations are not meant to serve as quantitative numerical simulations, but instead offer qualitative insight into the underlying mechanisms responsible for the observations. Lattice sites are occupied by isotropic 5 nm gold spherical dipoles described by a bulk Drude dielectric model with high-frequency dielectric constant $\varepsilon_\infty= 9.7,$ damping rate $\hbar\gamma=0.073 \text{ eV},$ and plasma frequency $\hbar\omega_p=8.96 \text{ eV}$. The collective excitations of the NCSLs are calculated by solving the direct-space coupled-dipole equations under either optical plane wave \autocite{zou2004,markel1993,cherqui2019} or STEM electron beam drive \autocite{bigelow2012,mirsaleh2012,bourgeois2022,rossi2024}, where each spherical dipole is coupled to all other dipoles in the finite patch via fully retarded dipole-dipole interactions.

The results of the coupled-dipole model are shown in Fig. \ref{fig:6}. The impact parameter $\mathbf{b} = (0, d/2)$ nm chosen to represent the EEL spectra is shown, denoted by an orange marker, in Fig. \ref{fig:6}a. The impact parameter is located in the vicinity of the patch center to minimize the influence of edge effects. Consistent with the results of the experiment, peaks of the calculated EEL spectra are blue-shifted relative to the optical extinction spectra, shown in Fig. \ref{fig:6}a. Additionally, peaks of the calculated bilayer spectra are blue-shifted relative to the monolayer spectra, which persist in both EEL and optical extinction spectra.

Under normally incident optical illumination, the many-dipole coupling can be reduced to an effective two-dipole model of plasmon hybridization, illustrated in Fig.~\ref{fig:6}b,\autocite{west2020,olafsson2021,nelsonquillin2016} which is qualitatively consistent with experimental and calculated spectra. The lowest energy collective plasmonic excitation of each individual hexagonal patch is an in-plane dipolar mode with the largest dipole moment; this mode is the only mode accessible to plane-wave optical excitation. Upon introduction of a second layer, the two in-plane collective dipolar modes of the monolayers hybridize into a higher-energy in-phase collective hybridized mode, which is the origin of the blue shift observed in both the experiment and coupled dipole calculations. This mode is accessible to both the optical plane wave and the electron beam, as indicated by the induced polarizations illustrated in Fig.~\ref{fig:6}c, and its spectral position is indicated by the red line in Fig.~\ref{fig:6}a. All other inter-layer hybridizations are inaccessible under normally incident plane-wave optical illumination, and thus the bilayer optical spectrum is blue-shifted relative to the monolayer spectrum. In the EEL spectra shown in Fig.~\ref{fig:6}a, the additional peaks at 2.64~eV (blue) and 2.70~eV (green) correspond to modes with out-of-plane polarizations which are accessible to the electron beam probe, on account of the three-dimensional polarization content of the electron's field. Due to their relatively broad linewidths, the presence of these additional spectral features causes the EEL spectrum to appear blue-shifted relative to the optical spectrum, consistent with experimental observations. The additional blue-shift in the bilayer EEL spectrum relative to the monolayer EEL spectrum is attributed to inter-layer hybridization. Although these hybridized modes are not readily described by an effective two-dipole model, the observed blue-shift is captured within the presented numerical solution to the coupled-dipole model.

The blue shift observed in the EEL spectra near twist centers in the twisted bilayer region arises from the decreased average distance to nearest neighbors in the relaxed, twisted configuration as compared to the AA stacked configuration. To understand the observed twist angle dependence on the EEL spectra near twist centers, the surrounding region is modeled by a small hexagonal lattice patch of 14 sites in the region of a twist center (Fig. S10), explicitly treating only the immediate region surrounding the twist center. Next, we model a rigid bilayer by keeping the vertical height of the lattice sites in the top layer constant. To model the relaxation that occurs as the top layer is twisted, lattice sites in the top monolayer shift downwards as it rotates, until the center-to-center distance between NCs in the top and bottom layers is again equal to the lattice spacing, filling the gap created by rotation (Fig. S10). When the top monolayer is treated as rigid, the EEL spectral peak positions are the same as those of the AA stacked configuration. This result suggests that although twisting moves each lattice site in the top monolayer away from its vertically aligned partner in the bottom monolayer, it simultaneously moves the site closer to neighboring sites in the bottom monolayer. As a result, the two effects cancel and the spectrum is left largely unchanged. When the top monolayer is allowed to relax as the lattice twists, a noticeable blue-shift occurs in the EEL spectrum, since the overall distance between lattice sites decreases, increasing the coupling strength and further blue-shifting the spectrum. The above analysis of lattice relaxation highlights that twist angle alone is not sufficient to explain the local blue-shifts observed around twist centers, and must be coupled to changes in interlayer spacing consistent with a non-rigid NCSL bilayer.

\section{Conclusion}
In this study, we fabricated high-quality Au NCSL monolayers through liquid-air self-assembly. Inspired by twisted vdW 2D materials, the NCSL monolayers are stacked to form moir\'e NCSLs. Our fabrication method provides direct insight into ways to create complex plasmonic metamaterials with rich optical responses. We demonstrated that low-loss EELS in a monochromated, aberration-corrected STEM provides ultra-high spatial and energy resolution for probing lattice plasmonic excitations in structurally heterogeneous NC assemblies, where conventional optical spectroscopy techniques exhibit fundamental limitations of ensemble averaging. STEM characterized the detailed stacking between the NCSL monolayers, enabling direct correlation between local nanocrystal arrangement and plasmonic response. EELS shows a blue shift of the plasmonic excitation from monolayer, bilayer, to twisted bilayer. Moreover, the EELS measurements exhibited an overall blue shift relative to optical microspectroscopy data. Model calculations using the method of coupled dipoles explained this overall blue shift by proving that EELS provides complementary information to optical microspectroscopy. The three-dimensional polarization structure of the electron beam excites higher-energy hybridized plasmonic lattice modes with out-of-plane polarizations, which are inaccessible under normally incident optical illumination. The high spatial resolution of STEM-EELS also demonstrated that, unlike the twisted vdW 2D materials, the twist-induced blue shift in plasmonic response of our NCSL system is highly localized because of the non-rigid nature of the monolayers.

Looking forward, our findings motivate several future directions. First, because the localization of twist-induced shifts arises from relaxation of non-rigid monolayers, stiffening the NCSLs through ligand cross-linking, inorganic ligand exchange, or rigid dielectric spacers should allow twist-dependent coupling to extend across entire moiré domains, enabling long-range twist-tunable responses detectable even by far-field optics. Second, deterministic stacking of free-standing NCSL membranes could provide controlled twist angles over large areas, permitting a systematic study of angle dependence from small twists to the 30° quasicrystalline configuration. Conversely, the strongly localized response at twist centers offers a route to positionally defined plasmonic hotspots for surface-enhanced spectroscopies. Finally, heterostructured stacks combining plasmonic NCSLs with excitonic nanoplatelet or perovskite layers could introduce moiré-modulated plasmon–exciton coupling, expanding the design space of programmable metamaterials.

\section{Methods}

\subsection{Synthesis of moir\'e Au NCSLs}

Spherical Au NCs with 5 nm diameter\autocite{fafarman2013,yang2018} were synthesized and dispersed in hexane at 2~mg/ml for self-assembly. A well-established liquid-air interface method was applied to form the monolayer NCSLs.\autocite{dong2010} First, 1.8~ml diethylene glycols (DEG) was added as the subphase into a $1.5 \times 1.5 \times 1$~cm\textsuperscript{3} Teflon trough. Then, 20~\textmu L of Au NCs in hexane was added and the trough was covered with a glass slide to allow the slow solvent evaporation. A TEM grid was used to dip the surface and transfer the monolayer NCSLs. This process was repeated to deposit the second layer onto the same TEM grid.\autocite{dong2011}

\subsection{Optical microspectroscopy of moir\'e NCSLs}

Absorption spectra of individual NCSL domains, along with optical images, were collected using a 308 PV microscope spectrophotometer system (Craic Technologies) integrated with an Olympus BX51 upright optical microscope. TEM grids containing the NCSLs were mounted on a glass slide. The samples were illuminated from below with unpolarized white light generated by a 100~W tungsten-halogen lamp. Transmitted light was captured using a 100$\times$ objective lens (LMPlanFL, NA 0.80), then focused onto the spectrophotometer's entrance aperture and recorded with a CCD detector. A region of $3.1 \times 3.1$~\textmu m can be captured each time.

\subsection{Acquisition of EELS data}

Our EELS study was conducted using a Nion UltraSTEM 100MC `HERMES' at the SuperSTEM Laboratory, Daresbury, United Kingdom (UK). The dedicated STEM uses a 100~kV cold FEG emitter with 0.3~eV native energy spread and has post-specimen optics which optimize 5\textsuperscript{th} order aberrations for high spatial resolution ($<0.7$~\AA\ at 100~kV). When combined with a Dectris ELA direct electron detector, this instrument has a demonstrated energy resolution of 5~meV.\autocite{hage2019} Here the microscope was operated at 60~kV acceleration voltage, with a probe convergence semi-angle of 30~mrad and a probe current of approximately 35~pA before monochromation. The monochromator slit was adjusted to improve the energy resolution to $<43$~meV (as estimated by the full-width at half-maximum of the zero-loss peak in vacuum), much narrower than all the spectral features we are observing and thus providing an optimal compromise of resolution and beam current. The energy dispersion of 10~meV per channel (or pixel) for the detector provides good precision for fitting features, yielding a sufficient number of pixels per spectral feature. The high-angle annular-dark-field detector angular range is calibrated as 90--185~mrad. Spectra were acquired using a collection semi-angle of 44~mrad.

To maintain the spatial resolution of the HAADF-STEM images, a pixel size of 1.2~nm was used. To avoid adventitious carbon build-up due to interaction between the beam and the ligands from the film's synthesis releasing carbon-based contamination in the otherwise ultra-high-vacuum environment of the microscope's column, and the resulting interference of the plasmonic peak of carbon with that of gold, a pixel dwell time of 0.03~s was chosen.

\subsection{Mapping thickness profile from EELS data with HyperSpy}

The log-ratio method\autocite{egerton2011} gives the thickness normalized by the electron mean free path for a particular sample and energy of the electron. This method for estimating the thickness from EELS is inaccurate for our study because we are only partially acquiring the low loss data (0--10~eV); further contributions from bulk plasmons at higher energy losses would need to be included to improve the accuracy. So, we implemented a different method where the change in thickness ($\Delta t$) is proportional to the change in intensity (area under the curve) of the zero-loss peak $\Delta I_0$. Such a method is also likely inaccurate. In our case, it works well enough because we have well defined integer numbers of layers, each with the same thickness, and channeling / plural scattering may have not yet made things too complicated. So, finding a scaling with ZLP intensity within the same dataset offers an ad hoc way to estimating local thickness jumps. Fig.~S9 shows the comparison between the thickness maps from the aforementioned methods. We can see the log-ratio method results in over-estimation of thickness in our case.

\subsection{Lattice constants used in the coupled-dipole model}
The experimental HAADF-STEM image of a self-assembled monolayer was used to calculate the lattice parameters used in the coupled dipole modeling. Though 5 nm Au NCs were used in our model, the ligands increase the lattice parameter $d$ to 6.84 nm. For the optical calculations, the bilayer regions included both AA and AB stacking regions, as well as moir\'e regions with a 19$^\circ$ relative twist angle. AA stacked regions were modeled using an inter-layer spacing equal to the dipole spacing, with AB stacked regions having an interlayer spacing of $\sqrt{2/3}\,d$, consistent with hexagonal close packing. The moir\'e interlayer spacing was taken to be equal to the lattice spacing, assuming rigid, incommensurate stacking. For twisted bilayers which allow for relaxation, the interlayer spacing of lattice sites around the twist center is set to $\cos(\varphi)d$, where $\varphi$ is the twist angle.

\section*{Data Availability}

Experimental data are available from the authors upon reasonable request. Scripts and notebooks for processing the EELS data are available at 

\url{https://github.com/swarnendudas2001/Hyperspy-EELS-mapping}.

\begin{acknowledgement}
This research was supported in Phase 2 by the NSF Centers for Chemical Innovation Program Grant (NSF CHE-2503933) from the Division of Chemistry. We also acknowledge use of the SuperSTEM, a United Kingdom national facility for Advanced Electron Microscopy Funded by EPSRC (EP/W021080/1).
\end{acknowledgement}

\printbibliography

\end{document}